\documentclass[12pt]{spieman}  
\usepackage{amsmath,amsfonts,amssymb}
\usepackage{graphicx}
\usepackage{setspace}
\usepackage{tocloft}
\usepackage{subfig}
\usepackage{booktabs}
\usepackage{lineno}
\usepackage{soul}
\usepackage{color}

\usepackage[bordercolor=white,backgroundcolor=gray!30,linecolor=black,colorinlistoftodos]{todonotes}

\title{A Light-Weighted Convolutional Neural Network for Bitemporal SAR Image Change Detection}

\author[a]{Rongfang Wang}
\author[a]{Fan Ding}
\author[a]{Licheng Jiao}
\author[a,*]{Jia-Wei Chen}
\author[a]{Bo Liu}
\author[a]{Wenping Ma}
\author[b]{Mi Wang}

\affil[a]{Key Laboratory of Intelligent Perception and Image Understanding of Ministry of Education, School of Artificial Intelligence, Xidian University, Xi'an, China, 710071}
\affil[b]{State Key Laboratory of Information Engineering in Surveying, Mapping and Remote Sensing, Wuhan University, Wuhan 430079, China.}

\cftpagenumbersoff{figure}
\cftpagenumbersoff{table} 
\graphicspath{{figures/}}

\begin{document} 
\maketitle

\begin{abstract}
Recently, many Convolution Neural Networks (CNN) have been successfully employed in bitemporal SAR image change detection. However, most of the existing networks are too heavy and occupy a large volume of memory for storage and calculation. Motivated by this, in this paper, we propose a lightweight neural network to reduce the computational and spatial complexity and facilitate the change detection on an edge device. 
In the proposed network, we replace normal convolutional layers with bottleneck layers that keep the same number of channels between input and output. Next, we employ dilated convolutional kernels with a few non-zero entries that reduce the running time in convolutional operators. {Comparing with the conventional convolutional neural network, our light-weighted neural network will be more efficient with fewer parameters. We verify our light-weighted neural network on four sets of bitemporal SAR images. The experimental results show that the proposed network can obtain better performance than the conventional CNN and has better model generalization, especially on the challenging datasets with complex scenes.}
\end{abstract}


{\noindent \footnotesize\textbf{*}Corresponding Author: Jia-Wei Chen~ \linkable{jawaechan@gmail.com} }

\begin{spacing}{1.5}   

\section{Introduction}
Over past decades, much attentions focus on multitemporal Synthetic Aperture Radar (SAR) image change detection, since a SAR is capable of working in all-time and all-weather without the influence of extremely bad weather and the cloud. 
In a past decade, most traditional SAR image change detection methods are developed how to extract changed areas from a difference image (DI), which suppose to include the information of changed regions. The DI calculated by the log-ratio (LR) \cite{bazi2006automatic} is usually subject to the speckle and it is challenging to extract the accurate and clear information on the changed region. To tackle this issue, sparse learning\cite{wang2016sar} was recently proposed learning robust features from the noisy DI. Wang et al.\cite{wang2019can} analyzed the affects of the SAR image speckle on change detection and proposed a sparse learning-based method for SAR image change detection.

Recently, deep neural networks have been successfully employed to computer vision and remote sensing image analysis due to its ability to exploiting essential and robust structural features on categories of objects. It also has been introduced into the field of change detection. Gong et al\cite{Gong2017Change} proposed a deep neural network for SAR image change detection. Gao et al. \cite{gao2016automatic} proposed a simple convolutional neural network, known as PCA-Net, exploring robust features on the changed regions from the noisy DI. However, the performance of these two unsupervised methods are limited without the correct guidance. To tackle this issue, Wang et al. \cite{wang2018imbalanced} proposed a supervised PCA-Net to improve the performance by carefully collecting typical training samples, which obtains state-of-arts performance of bitemporal SAR image change detection. However, this two-layer convolutional neural network is low efficient, since the convolutional kernels are trained or generated by a Principle Component Analysis (PCA) decomposition. Recently, Li et al. \cite{li2019deep} proposed a convolutional neural network (CNN) for SAR image change detection based on both unsupervised and supervised learning. Zhao et al. \cite{zhao2017novel} proposed a bitemporal PolSAR image change detection by a joint classification and a similarity measurement.  Currently, it is still an open problem to extract the changed regions from the noisy DI. 
{Nowadays, a large volume of SAR images are acquired by satellites and it is imperative to develop an efficient model that can produce promising results of SAR image interpretation. Most above networks are too heavy and cost much computational burden. It is strongly required to develop a lightweight convolutional neural network.}

 {Recently, several lite networks are proposed to improve the inference efficiency. Howard et al. and Sandler et al. proposed two lite networks MobileNetV1 \cite{howard2017mobilenets} and MobileNetV2 \cite{sandler2018mobilenetv2} for visual category. Recently, Howard et al. proposed to search for MobileNetV3 \cite{howard2019searching}. Tan et al. \cite{tan2019efficientnet} proposed an efficient network for visual category. These lite networks have been extensively employed to visual category and the experimental results show that they can achieve comparable performance with heavy networks, but with low latency and network capacity. It can be potentially performed on edge devices with low power. Most recently, Chen et al.\cite{chen2020a} proposed a lightweight multiscale spatial pooling network for bitemporal SAR image change detection. 

Following the idea of the lightweight neural network, in this letter, we focus on the application of lite networks in SAR image change detection. To achieve this, we propose a lite CNN for SAR image change detection. In the proposed network, bottleneck layers are introduced to reduce the number of output channel. Furthermore, the dilated convolutional layers\cite{li2018csrnet} are introduced to enlarge receptive field with a few of non-zero entries in the kernel, which reduces the number of network parameters. We verify the proposed network by comparing other conventional CNN. Compared with the lightweight network in \cite{chen2020a}, the proposed network is more robust with the residual and bottleneck structure. Experimental results on four sets of bitemporal SAR image show that our proposed method obtain comparable performance with CNN, while being much more efficient than CNNs. }

The rest of paper will be organized as follows. We will introduce our proposed method in Section 2. Then the proposed method will be verified on four datasets in Section 3. Finally, we will draw a conclusion in Section 4.

\section{Proposed Method}
{Given bitemporal SAR images ${\bf I}_1$ and ${\bf I}_2$, the DI can be generated as follows}
 \begin{equation}\label{di}
 {\bf I}_{DI} = {\bf I}_1 \ominus {\bf I}_2
 \end{equation}
 {where $\ominus$ denote the difference operator.  However, most existing difference operator is subject to the speckle. 
Then we will propose a lightweight convolutional neural network to exploit the changed regions from the noisy DI.}

\begin{figure}[!htb]
\centering
\includegraphics[width=\textwidth]{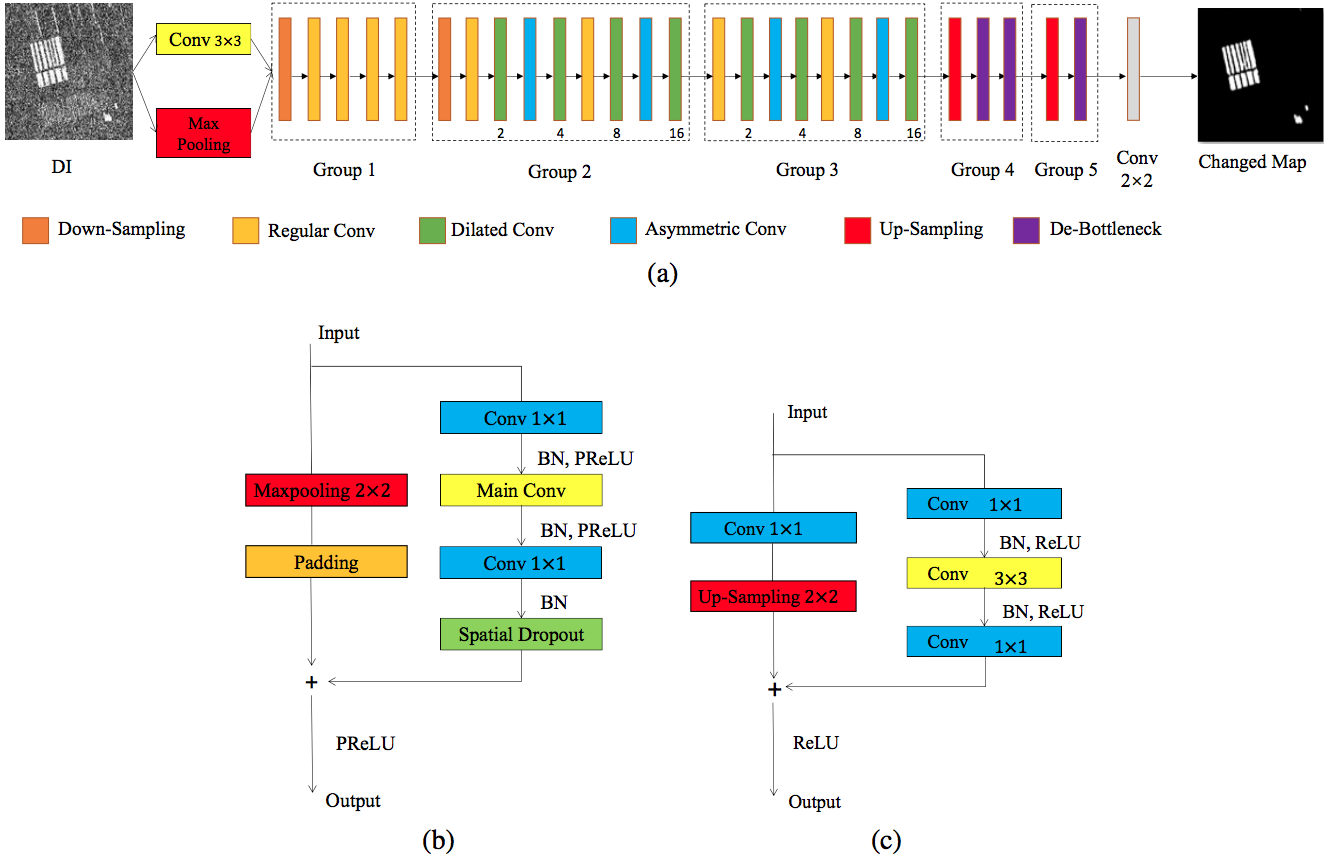}
\caption{The framework of the proposed network.(a)Network Architecture. (b) Bottleneck for encoder. (c)Bottleneck for decoder.}
\label{framework}
\end{figure}

\begin{table}[!htbp]
\centering
\caption{The tensors of all the layers.}
\label{tab:network}
\begin{tabular}{cccccccc}
\toprule
 Layer Name & Tensor Size & & Layer Name & Tensor Size & & Layer Name & Tensor Size   \\
  \hline
   \multicolumn{8}{c}{Initial Block} \\
Input& (32,32,1) \\ 
 Conv&(16,16,13) & & 
 Max-pooling & (16,16,1) && 
 Concatenation & (16,16,14) \\
 \hline
  \multicolumn{8}{c}{Group 1} \\
BottleNeck 1.0&(8,8,64) & &
BottleNeck 1.1&(8,8,64) & &
BottleNeck 1.2&(8,8,64)\\
BottleNeck 1.3&(8,8,64) & &
BottleNeck 1.4&(8,8,64) \\
\hline
 \multicolumn{8}{c}{Group 2} \\
BottleNeck 2.0&(4,4,128) & &
BottleNeck 2.1&(4,4,128) & &
BottleNeck 2.2&(4,4,128)\\
BottleNeck 2.3&(4,4,128) & &
BottleNeck 2.4&(4,4,128) & &
BottleNeck 2.5&(4,4,128) \\
BottleNeck 2.6&(4,4,128) & &
BottleNeck 2.7&(4,4,128)& &
BottleNeck 2.8&(4,4,128) \\
\hline
 \multicolumn{8}{c}{Group 3} \\
BottleNeck 3.0&(4,4,128) & &
BottleNeck 3.1&(4,4,128) & &
BottleNeck 3.2&(4,4,128)\\
BottleNeck 3.3&(4,4,128) & &
BottleNeck 3.4&(4,4,128) & &
BottleNeck 3.5&(4,4,128)\\
BottleNeck 3.6&(4,4,128) & &
BottleNeck 3.7&(4,4,128)\\
\hline
\multicolumn{8}{c}{Group 4} \\
BottleNeck 4.0&(8,8,64) & &
BottleNeck 4.1&(8,8,64) & &
BottleNeck 4.2&(8,8,64)\\
\hline
\multicolumn{8}{c}{Group 5} \\
BottleNeck 5.0&(16,16,16) & &
BottleNeck 5.1&(16,16,16) \\
\hline
\multicolumn{8}{c}{Output} \\
Conv&(32,32,2) \\
  \bottomrule
\end{tabular}
\end{table}

The whole framework of the proposed network can illustrated in Fig.\ref{framework}(a). It is shown that the network consists of five groups of bottleneck layers \cite{lin2013network} with an 1$\times$1 kernels, illustrated by variety of colorful bars, among which the former three ones work as the encoder, and the latter two ones as the decoder. {The tensors of all layers are listed in Table \ref{tab:network}.}

In the forward process, the network takes a patch of DI as the input. Firstly, the input data go through a normal convolutional layer and a max-pooling (MP) layer, respectively and then the outputs are concatenated. Next, the contact activations go through the decoder with three groups bottleneck layers. The essential structure of a bottleneck of encoder can be illustrated in Fig.\ref{framework}(b). It is shown that a bottleneck layer is constructed by a small residual block, including a maxpooling path and a convolutional path. More specifically, the convolutional path consists of two 1$\times$1 convolutions and one main convolution. The main convolution will vary with the various function of the bottleneck. It can be a normal convolution, a dilated convolution \cite{li2018csrnet} or an asymmetrical convolution\cite{szegedy2016rethinking}. {The tensors inside the encoding bottleneck are listed in Table \ref{tab:bo_en}.}   
\begin{table}[!htbp]
\centering
\caption{The tensors inside an encoding bottleneck layer.}
\label{tab:bo_en}
\begin{tabular}{ccccc}
\toprule
 Layer Name & Tensor Size & & Layer Name & Tensor Size   \\
  \hline
  Input& (16,16,14) \\
  \hline
  \multicolumn{5}{c}{Branch 1} \\
 Conv1& (8,8,16) &&
Conv2&(8,8,16) \\
Conv3&(8,8,64) &&
Dropout&(8,8,64) \\
\hline
\multicolumn{5}{c}{Branch 2} \\
Max-pooling&(8,8,14) & &
Padding&(8,8,64)\\
\multicolumn{5}{c}{Output} \\
\hline
Addition & (8,8,64)\\
  \bottomrule
\end{tabular}
\end{table}
\begin{table}[!htbp]
\centering
\caption{The tensors inside a decoding bottleneck layer.}
\label{tab:bo_de}
\begin{tabular}{ccccc}
\toprule
 Layer Name & Tensor Size & & Layer Name & Tensor Size   \\
  \hline
  Input& (4,4,128) \\
  \hline
  \multicolumn{5}{c}{Branch 1} \\
Conv1& (4,4,16) &&
Conv2&(8,8,16) \\
Conv3&(8,8,64) \\
\hline
\multicolumn{5}{c}{Branch 2} \\
Conv&(4,4,64) & &
Up-Sampling&(8,8,64)\\
\multicolumn{5}{c}{Output} \\
\hline
Addition & (8,8,64)\\
  \bottomrule
\end{tabular}
\end{table}

 The first group consist a down-sampling bottleneck and four normal convolutional bottleneck layers. In the normal bottleneck layers, the main convolution component is default set as the main normal convolutional layer. Especially, in the down-sampling bottleneck, the main convolutional component is set as a normal convolution kernel with 3$\times$3 and the 1$\times$1 convolution component is replaced by a 2$\times$2 one. In the next two groups, to exploit the spatial context, we insert the bottleneck layers with asymmetrical and dilated convolution layers with various kernel sizes among the normal bottleneck, where the kernel sizes are set 2, 4, 8 and 16, respectively, as shown the digits below the green bars. In these bottleneck layers, the main convolutional layers are replaced by the dilated convolutional layers, where the kernels are sparse and most entries are zeros. Furthermore, the bottleneck layers with asymmetric convolution layers are also insert between the dilated and normal convolutional layer, illustrated by the blue bars. After encoding, the sizes of feature maps decrease as the one fourth as the original image. 

In the decoding part, the context information collected by the encoder will be propagated to the pixel level. To achieve this, inspired by the idea of U-Net \cite{ronneberger2015u}, we schedule a upsampling layer and two bottleneck layers. The essential structure of bottleneck for decoder can be illustrated in Fig.\ref{framework}(c). Similar to the bottleneck for the encoder, the bottleneck for the decoder contains a pooling path and a convolution path. The former includes a maxpooling layer and a 1$\times$1 convolution, while the latter includes two 1$\times$1 convolutional layers and a 3$\times$3 convolutional layer. Especially, when the bottleneck layer is used for upsampling, the 3$\times$3 convolution component, illustrated by the yellow bar, will be replaced by a 3$\times$3 transpose convolution \cite{dumoulin2016guide}. Then the bottleneck will do the 2x upsampling. Trough two groups of decoding bottleneck layers, the feature map will be recover the same size as the input image. {The tensors inside the decoding bottleneck are listed in Table \ref{tab:bo_de}.}  Finally, we put a 2$\times$2 convolutional layer to get the probability map of two categories. 

\begin{figure}[!htb]
\centering
\includegraphics[width=\textwidth]{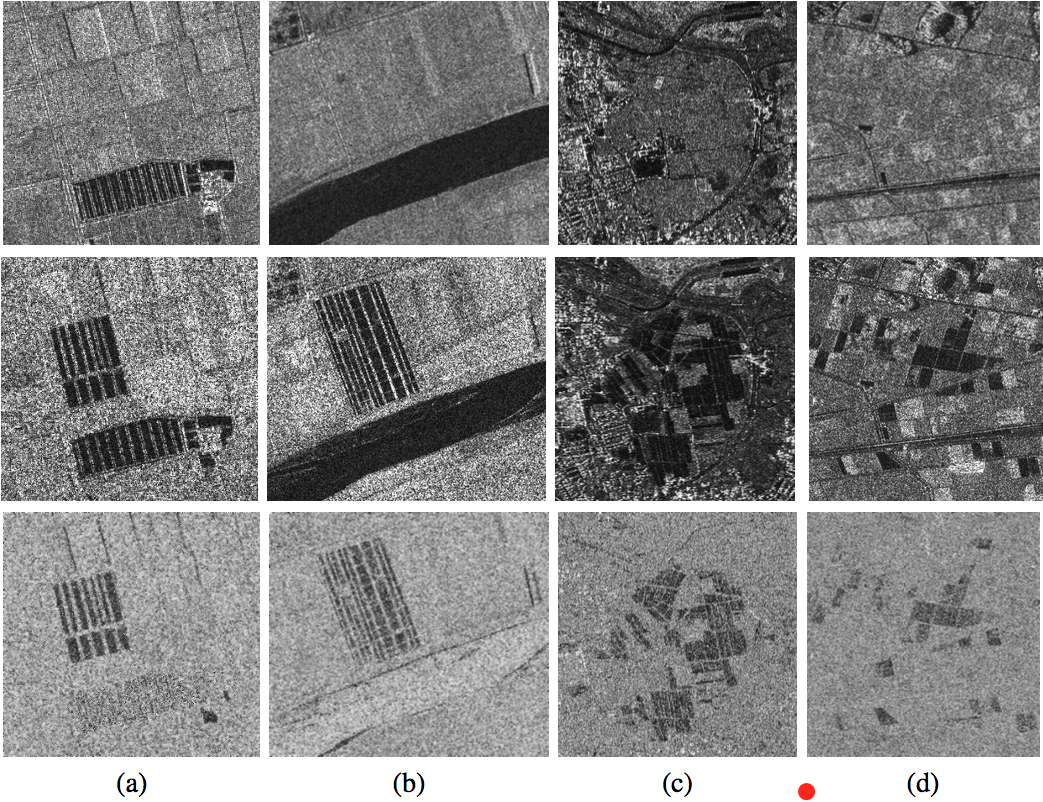}
\caption{Four sets of bitemporal SAR images. The first two rows are bitemporal images and the last row is the DIs. (a) YR-A. (b) YR-B. (c) Sendai-A. (d) Sendai-B. }
\label{fig2}
\end{figure}

\section{Experimental Results}
\label{sec:experiment}
\subsection{Experiment Datasets}
In this paper, the proposed method is verified on four sets of bitemporal SAR images. Two scenes (YR-A and YR-B) are from bitemporal Yellow River SAR images \cite{Gong2017Change} acquired by the Radarsat-2 satellite in 2008 and 2009, respectively. Their image sizes are 306 $\times$ 291 and 400 $\times$ 350, respectively. Other two are parts of TerraSAR-X images acquired prior to (on Oct. 20, 2010) and after (on May 6, 2011) the Sendai earthquake in Japan \cite{Cui2016A}. Their sizes (Sendai-A and Sendai-B) are 590 $\times$687 and 689 $\times$ 734, respectively. These four datasets are shown in Fig.\ref{fig2}). These four datasets are quite challenging, such as the linear-shape changed regions in YR-B dataset and complex scene in both Sendai-A and Sendai-B datasets. 

\begin{table}[!htbp]
\centering
\caption{The number of training samples.}
\label{tab:nos}
\begin{tabular}{ccccc}
\toprule
  Dataset & YR-A & YR-B & Sendai-A & Sendai-B   \\
  \hline
  No. Samples& 3596 & 6205 & 15375 & 20294 \\
  \bottomrule
\end{tabular}
\end{table}
\begin{figure}[!htb]
\centering
\includegraphics[width=0.6\textwidth]{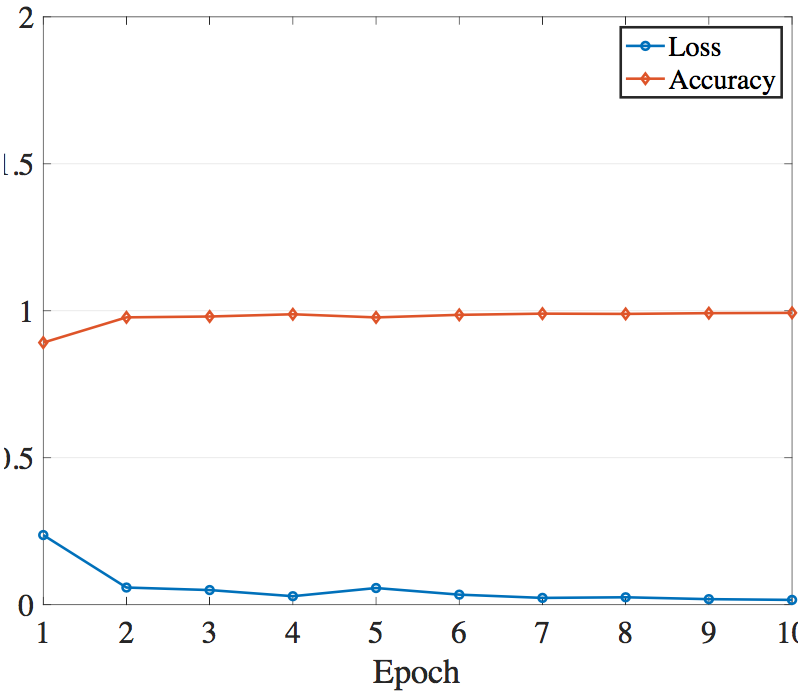}
\caption{The variations of loss and accuracies.}
\label{fig:loss}
\end{figure}

\subsection{Implementations}
{
We have introduced a lite CNN for change detection for bitemporal SAR images. We first generate the DI by the Eq.(\ref{di}) and the difference operator is implemented by the neighborhood-based LR operator \cite{gong2011neighborhood}. 
To train the network, we collect a training dataset according to the method in \cite{wang2018imbalanced}. The numbers of training samples for each dataset are listed in Table \ref{tab:nos}.

More specifically, the patchsize of each sample is set as 32 $\times$ 32 and 8 samples are fed in each training step. Additionally, the network is trained by an end-to-end back-propagation manner and the loss is set as {the binary entropy function defined in \cite{sadowski2016notes}}. 
The training is optimized by the Adam algorithm \cite{kingma2014adam} in the training stage, where the initial learning rate is set as 0.005. 
The training is performed on the PyTorch platform built on the Ubuntu 16.04 installed in a PC with a 16 GB DDR memory and an NVIDIA TITAN Xp Graphics Processing Unit of 11 GB memory. The training process will converge at around 15 epochs. We show the variations of loss values and accuracies of training process in Fig.\ref{fig:loss}.
}
\subsection{Comparison Experiments}
To verify the benefits of the proposed method, it is compared with the unsupervised PCA-Net (U-PCA-Net)\cite{gao2016automatic}, the supervised PCA-Net (S-PCA-Net) \cite{wang2018imbalanced} which achieves the state-of-arts performance on SAR image change detection. We also compare the proposed method with the deep neural network (DNN) method \cite{Gong2017Change} and CNN \cite{li2019deep}. Among these methods, DNN and U-PCA-Net are unsupervised methods,while S-PCA-Net, CNN and the proposed method are supervised ones. 

The performance of the compared methods is evaluated by probabilistic Missed Alarm (pMA), probabilistic False Alarm (pFA) and kappa coefficient, where pFA (pMA) are calculated by the ratios between FA (MA) and the number of Non-Changed pixels (NC)\cite{wang2019can}. 

\begin{figure}[!thb]
\centering
\includegraphics[width=\textwidth]{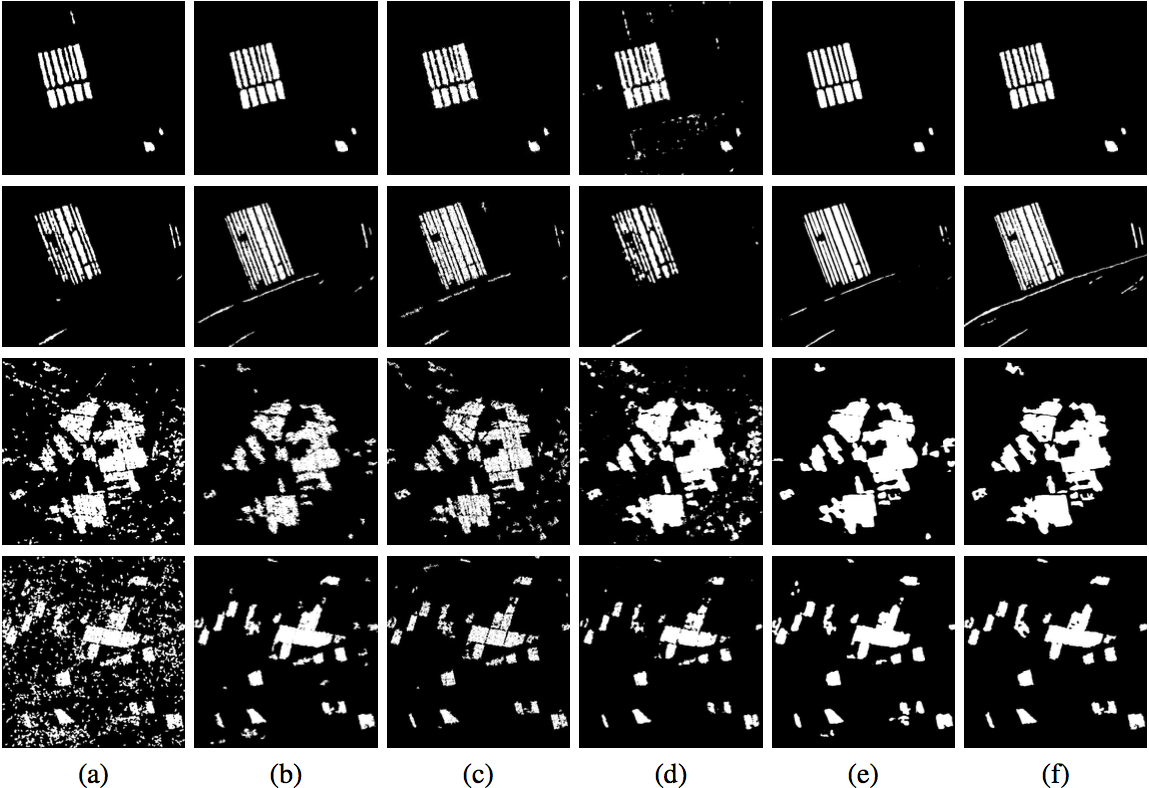}
\caption{The visual comparison results. (a)U-PCA-Net. (b)S-PCA-Net. (c)DNN. (d) CNN. (e)Lite CNN. (f)Reference. }
\label{fig3}
\end{figure}
\begin{figure}[!htb]
\centering
\includegraphics[width=\textwidth]{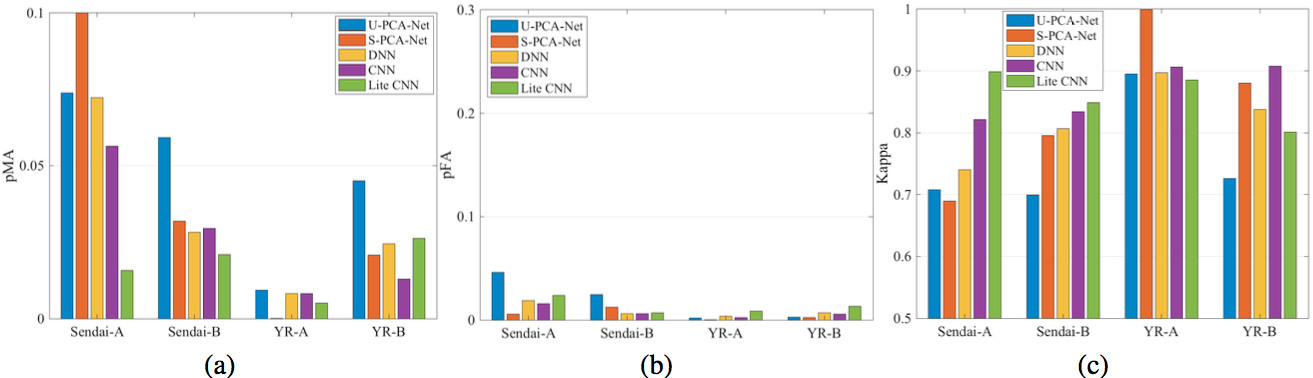}
\caption{The quantitative evaluations of compared methods.(a) MA. (b)FA. (c) Kappa. }
\label{eva}
\end{figure}
\subsection{Experiment Results on Individual Dataset Change Detection}
In this experiment, for the supervised learning methods, we collect the training samples from an individual dataset, which covers 30\% areas of the whole image frame. The rest part is employed for testing.  

The visual comparison results are shown in Fig.\ref{fig3}. It is shown that for the YR-A dataset, S-PCA-Net, DNN and Lite CNN get less noisy but more completed changed regions. Lite CNN can get more clear boundary of changed regions than DNN. For the YR-B dataset, Lite CNN can get more completed changed regions, especially the line at the bottom of the image. Other methods can not get the completed changed regions. For the Sendai-A dataset with complex scene, S-PCA-Net and the Lite CNN are less subject to the speckle and the background and get more clear changed regions, while other compared methods are subjected to the speckle and background and they are almost failed. Moreover, compared with S-PCA-Net, Lite CNN get better inner regional consistence.  For the Sendai-B dataset, Lite CNN gets more accurate changed regions than other methods.

Moreover, we show the quantitative evaluations in Fig.\ref{eva}. {It is shown in Fig.\ref{eva} (a) that the proposed Lite CNN gets a lower pMA on YR-A, Sendai-A and Sendai-B dataset, while the CNN gets a lower pMA on YR-B dataset. It is shown in Fig.\ref{eva} (b) that the proposed Lite CNN gets a lower pFA on Sendai-B dataset, while S-PCA-Net gets a lower pFA on YR-A, YR-B and Sendai-A dataset. }
It is shown in Fig.\ref{eva} (c) that on the YR-A dataset, S-PCA-Net gets the best kappa among all the methods, while the Lite CNN gets the comparable kappas with other methods, except S-PCA-Net. On the YR-B dataset, Lite CNN gets the comparable kappas with other methods. However, on both Sendai-A and Sendai-B datasets, the Lite CNN performs better than other methods in terms of pMA and kappas.  

\begin{figure*}[!hbt]
\centering
\includegraphics[width=\textwidth]{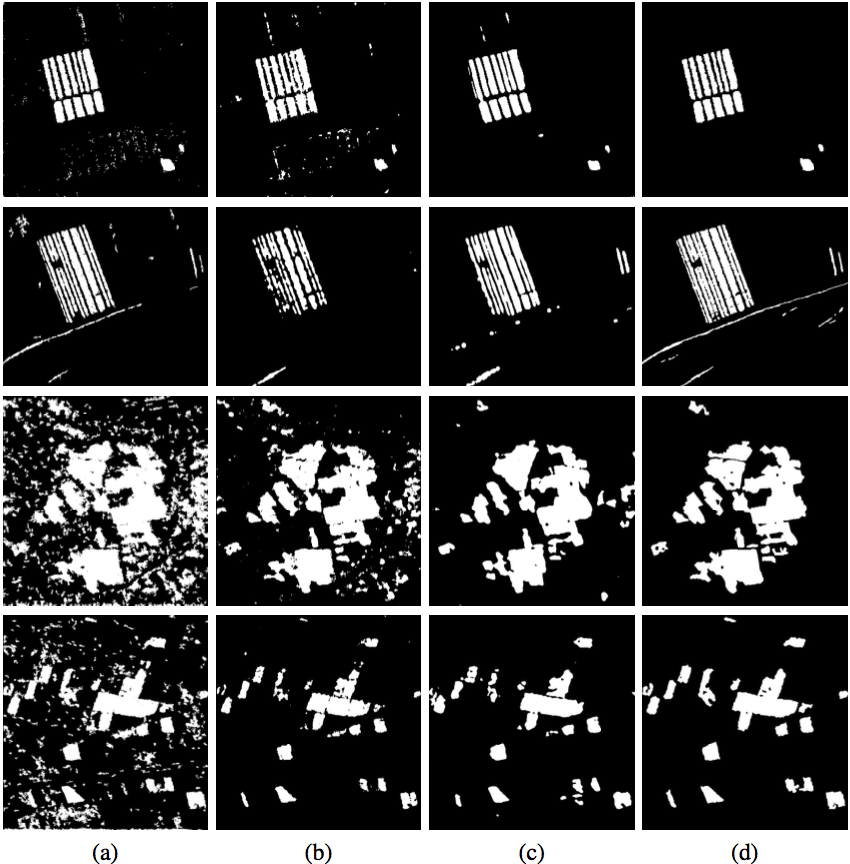}
\caption{The visual comparison results. (a) S-PCA-Net. (b) CNN. (c) Lite CNN. (d) Reference. }
\label{fig4}
\end{figure*}
\begin{figure*}[!htb]
\centering
\includegraphics[width=\textwidth]{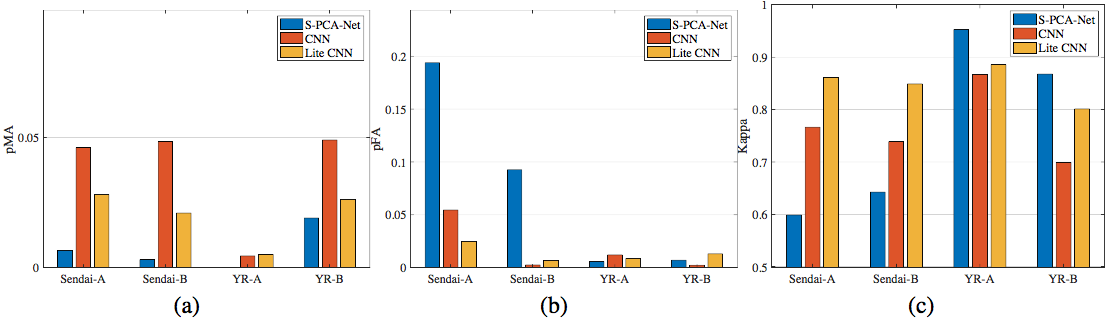}
\caption{The quantitative evaluations of compared methods.(a) MA. (b)FA. (c) Kappa.}
\label{eva2}
\end{figure*}
\subsection{Experiment Results on Cross-dataset Change Detection}
To further compare the proposed method with other supervised learning methods, S-PCA-Net and CNN, we perform the comparisons on the cross-dataset change detection, where the network trained on several datasets is applied to an unknown testing dataset. More specifically, to achieve this, this experiment is conducted through the leave-one-out manner, i.e. each dataset alternative is selected as the testing dataset and others as training datasets. 

The visual comparisons are shown in Fig.\ref{fig4}. It is shown that on the YR-A dataset Lite CNN gets more clear visual result with less noisy spots.  On the YR-B dataset, Lite CNN performs better than CNN, but not better than S-PCA-Net. There is many miss alarms in the results of Lite CNN. On both Sendai-A and Sendai-B datasets, Lite CNN gets better results than other two methods. 

{The quantitative evaluations in terms of pFA, pMA and Kappa are shown in Fig.\ref{eva2}. It is shown in Fig.\ref{eva2}(a) that the S-PCA-Net gets a lower pMA on all datasets. It is shown in Fig.\ref{eva2}(b) that the Lite CNN gets a lower pFA on the Sendai-A dataset, while S-PCA-Net gets a lower pFA on the YR-A dataset and CNN gets a lower pFA on the YR-B and Sendai-B dataset.} It is shown in Fig.\ref{eva2}(c) that Lite CNN performs better than CNN but comparable with S-PCA-Net on YR-A and YR-B datasets. However, Lite CNN shows great advantages over other two methods on Sendai-A and Sendai-B datasets. It indicates that the Lite CNN performs better than other two methods in model generalization, especially on challenging datasets with complex scenes.
\begin{table}[!htbp]
\centering
\caption{The training times of compared methods.}
\label{time}
\begin{tabular}{lccc}
\toprule
 Methods & S-PCA-Net & CNN& Lite CNN   \\
  \hline
Times& ~3 h & 30 mins. &15 mins. \\ 
  \bottomrule
\end{tabular}
\end{table}
\subsection{Discussion}
From the above comparisons, it is shown that Lite CNN can obtain comparable performance with other method on the YR-A and YR-B datasets. On the challenging datasets, e.g. Sendai-A and Sendai-B, Lite CNN outperforms other methods. Moreover, it has better ability to model generalization. Moreover, Lite CNN is more computationally efficient than S-PCA-Net and CNN. The training times of three supervised learning methods are compared in Table \ref{time}. It is shown that Lite CNN is easy to train and take less time than S-PCA-Net and CNN, while S-PCA-Net takes longer time, since the convolutional kernel is generated by the principle component analysis decomposition. 

Overall, Lite CNN can obtain comparable or even better performance than S-PCA-Net and CNN. It is also more computationally efficient than other two methods. It is expected that Lite CNN is more practical in change detection, especially for the requirement of real-time detection.

\section{Conclusion}
\label{sec:conlusion}
In this paper, we develop a lightweight convolutional neural network for bitemporal SAR image change detection. The proposed network consists of groups of bottlenecks layers which exploit the image feature. To verify the benefits of our proposed method, we compare it with several traditional neural networks and the comparisons are performed on fours sets of bitemporal SAR images. The experimental results show that our proposed method Lite CNN performs better than other two methods on cross-dataset change detection, especially when the scene is complex. Furthermore, Lite CNN is a lightweight network, which is more computationally efficient than CNN and S-PCA-Net. In the future, we will further optimize Lite CNN and make it more efficient on the edge device. 

 \section*{Acknowledgment}
 This work was supported by the State Key Program of National Natural Science of China (No. 61836009), the National Natural Science Foundation of China (No. 61701361, 61806154), the Major Research Plan of National Natural Science Foundation of China (No. 91838303), the Open Fund of Key Laboratory of Intelligent Perception and Image Understanding of Ministry of Education, Xidian University (Grant No. IPIU2019006), the Natural Science Basic Research Plan in Shaanxi Province of China (No.2018JM6083) and the Open Fund of State Laboratory of Information Engineering in Surveying, Mapping and Remote Sensing, Wuhan University (No. 17E02).

\bibliographystyle{spiejour}   
\bibliography{ref}

\end{spacing}
\end{document}